# Quantum fluctuations and life

P.C.W. Davies
Australian Centre for Astrobiology
Macquarie University, New South Wales, Australia 2109

## ABSTRACT

There have been many claims that quantum mechanics plays a key role in the origin and/or operation of biological organisms, beyond merely providing the basis for the shapes and sizes of biological molecules and their chemical affinities. These range from Schrödinger's suggestion that quantum fluctuations produce mutations, to Hameroff and Penrose's conjecture that quantum coherence in microtubules is linked to consciousness. I review some of these claims in this paper, and discuss the serious problem of decoherence. I advance some further conjectures about quantum information processing in bio-systems. Some possible experiments are suggested.

## 1. BACKGROUND

*"One can best feel in dealing with living things how primitive physics still is."*

Albert Einstein[1]

Fifty years ago physicists seemed on the verge of solving the riddle of life. Using quantum mechanics, great strides had been made elucidating the atomic and subatomic structure of matter, and the realization that life's secrets lay at the molecular level encouraged the expectation that quantum mechanics might play a key role in biology too. Certainly the founders of quantum mechanics thought so. Schrödinger's book *What is Life?* appealed to quantum ideas to bolster his claim that genetic information was stored on an "aperiodic crystal" structure[2]. He also speculated that quantum effects would lead to some sort of radically new behaviour that would distinguish the living from the nonliving. "We must be prepared to find a new kind of physical law prevailing on it," he wrote[3]. Bohr believed that the distinction between living and nonliving systems was fundamental, and actually a manifestation of his principle of complementarity[4].

In one sense these early pioneers were right. Quantum mechanics provides an explanation for the shapes of molecules, crucial to the templating functions of nucleic acids and the specificity of proteins. The Pauli exclusion principle ensures that atoms and molecules possess definite sizes, which in turn determines not just templating, but differential diffusion rates, membrane properties and many other important biological functions. Quantum mechanics also accounts for the strengths of molecular bonds that hold the machinery of life together and permit metabolism. But these examples are not quite what Schrödinger and Bohr were hinting at. They merely serve to determine the structure, stereochemistry and chemical properties of molecules, which may thereafter be treated as essentially classical. This leads to the ball-and-rod view of life, which is the one routinely adopted by biochemists and molecular biologists, according to which all essential biological functions may be understood in terms of the arrangements and rearrangements of classical molecular units of various shapes, sizes and stickiness.

But there are fundamental aspects of quantum mechanics that go beyond this description, such as:

a. Superpositions and various aspects of quantum phases, such as resonances.
b. Entanglement.
c. Tunneling.
d. Aspects of environmental interactions, such as the watchdog and inverse watchdog effects.
e. Abstract quantum properties such as supersymmetry.

Might it be that the qualitatively distinctive properties of life owe their origin to some aspect of "quantum weirdness"? In this paper I shall consider the possibility that one or more of the above properties may play a key role in the operation of biological or pre-biological processes.

## 2. CHEMISTRY AND INFORMATION

Although there is no agreed definition of life, all living organisms are information processors: they store a genetic database and replicate it, with occasional errors, thus providing the raw material for natural selection. The direction of information flow is bottom up: the form of the organism and its selective qualities can be traced back to molecular processes. The question then arises of whether, since this information flows from the quantum realm, any vestige of its quantum nature, other than its inherent randomness, is manifested.

Molecular biology is founded on a key dualism. Biological molecules serve two distinct roles: (i) specialized chemicals, (ii) informational molecules. This reflects the underlying dualism of phenotype/genotype. Using the analogy of computing, chemistry corresponds to hardware, information to software. A full understanding of the origin and function of life requires the elucidation of both the hardware and software aspects.

Studies of biogenesis have tended to focus on chemistry (i.e. hardware), by attempting to discover a chemical pathway from non-life to life. Though this work has provided important insights into how and where the basic building blocks of life might have formed, it has made little progress in the much bigger problem of how those building blocks were assembled into the specific and immensely elaborate organization associated with even the simplest autonomous organism[5].

Leaving open the questions of whether quantum mechanics may be important for understanding the emergence of chemical complexity in general, let me focus on the second aspect of the problem – the software, or informational content of bio-molecules. In recent years our understanding of the nature of information has undergone something of a revolution with the development of the subjects of quantum computation and quantum information processing[6]. Central to this enterprise is the replacement of the classical information "bit" by its quantum counterpart, the "qubit." As a quantum system evolves, input qubits transform into output qubits. Information may be processed by associating it with qubits; importantly, the processing efficiency is enhanced because quantum superposition and entanglement represent a type of computational parallelism. In some circumstances this enhancement factor can be exponential, implying a vast increase in computational speed and power over classical information processing. Several researchers have spotted the sweeping consequences that would follow from the discovery that living organisms might process information quantum mechanically, either at the bio-molecular level, or the cellular/neuronal level[5,7-10].

To be sure, biological systems are quintessential information processors. The informational molecules are RNA and DNA. Although quantum mechanics is crucial to explain the *structure* of these molecules, it is normally disregarded when it comes to their information processing role. That is, biological molecules are assumed to store and process *classical bits* rather than *qubits*. Here I shall entertain the possibility that, at least in some circumstances, that assumption may be wrong. We may then distinguish three possibilities of potential interest:

(i) Quantum mechanics played a significant role in the emergence of life from nonliving chemical systems in the first place, but ceased to be a significant factor when life got going.
(ii) Quantum information processing may have played a key role in the emergence of life, and a sporadic or subsidiary role in its subsequent development. There may be relics of ancient quantum information processing systems in extant organisms, just as there are biochemical remnants that give clues about ancient biological, or even pre-biological, processes.
(iii) Life started out as a classical complex system, but later evolved some form of quantum behavior as a refinement. For example, if biological systems can process information quantum mechanically, they would gain a distinct advantage in speed and power, so it might be expected that natural selection would discover and amplify such capabilities, if they are possible. This is an extension of the dictum that whatever technology humans invent, nature normally gets there first.

Although quantum information processing is more efficient than its classical counterpart, its scope for application is very severely circumscribed by the requirement of quantum coherence. This entails the preservation of delicate phase relationships between different components of the wave function over a long enough duration for information to be processed. Interactions between the quantum system and its environment will serve to decohere the wave function: the noise of the environment effectively scrambles the phases. Once decohered, a quantum system behaves in most respects as a classical system[11]. In particular, qubits degenerate into bits, and the enhanced information processing capabilities of the quantum system disappear. The central, and at present formidable, challenge in the race to develop a practical quantum computer is the need to reduce the effects of decoherence. Unless this can be done simply and efficiently, quantum computation will never become a practical reality. The decoherence rate depends on the nature and temperature of the environment and the strength with which it couples to the quantum system of interest[11,12], so the main burden in the development of quantum computation is to screen out the decohering environment as efficiently as possible, to reduce the temperature and to develop error correcting procedures to ameliorate the damage of environmental disruption. If quantum mechanics is to play a role in biological information processing, typical decoherence rates must not be greater than the biochemically significant reaction rates and/or error correcting processes must be deployed.

Simple models of decoherence have been much studied over the past twenty years. Typically, for a particle interacting with a heat bath at room temperature, exceedingly short decoherence times result[12]. Translated into the context of, say, a nucleotide in the environment of a cell at room temperature, decoherence times of less than $10^{-13}$ s appear typical. On the face of it, this is so rapid it is hard to imagine anything biologically significant emerging. However, there are several reasons why simplistic models of decoherence may not apply to biological systems. I defer a discussion of these reasons to Section 6.

## 3. GENERAL ARGUMENTS

A number of arguments on very general grounds have been deployed to establish when quantum processes may or may not be important in bio-systems, including the application of dimensional analysis and scaling laws. For example, Schrödinger reasoned, without having any detailed knowledge of the nature of DNA or heredity, that quantum mechanics was important in the stability of genetic information, and that quantum fluctuations might be the cause of some mutations[2]. He arrived at this conclusion on general energetic grounds. Wigner used an argument simply based on enumerating the dimensionality of Hilbert space that the replication of an organism could not be described as a unitary quantum process[13]. A modern variant of this argument appeals to the quantum no-cloning theorem[14], according to which a pure quantum state cannot be quantum mechanically replicated. These arguments, whilst interesting, are probably of limited relevance to real biology, where reproduction of genetic information does not entail the exact replication of an entire quantum state.

The quantum uncertainty principles imply certain limitations on the fidelity of all molecular processes, including biologically important ones. For example, the energy-time uncertainty relation sets a fundamental limit to the operation of all quantum clocks[15-17]. For a clock of mass $m$ and size $l$, Wigner found[15]

$$T < ml^2/\hbar. \qquad (1)$$

Curiously, for values of $m$ and $l$ of interest in molecular biology, $T$ also takes values of biological interest, suggesting that some biological systems operate at the threshold of quantum stability. To illustrate this point more precisely, let us suppose that some high-fidelity bio-molecular process requires choreography with a precision limited by Eq. (1). One example might be protein folding, which is well-known to be a major outstanding problem of theoretical biology[18]. To achieve its active conformation, a peptide chain of, say, $N$ amino acids must fold itself into a specific three-dimensional structure. The number of possible configurations is astronomical, and it is something of a mystery how the chaotically-moving chain "finds" the right configuration in such a short time (typically microseconds). If the average mass and length of an amino acid are $m_o$, and $a$ respectively, then Eq. (1) yields

$$T < m_o a^2 N^3/\hbar, \qquad (2)$$

suggesting a quantum scaling law for the maximum folding time of

$$T \propto N^3. \qquad (3)$$

The assumption that $l \equiv Na$ is the appropriate size factor in Eq. (1) may be justified for small proteins ($N = 80$ to 100) that fold in a one step, but larger proteins do not remain "strung out" for a large fraction of the folding process. Rather, they first fold into sub-domains. The opposite limit would be to replace $l$ by the diameter of the folded protein. Assuming it is roughly spherical, this would imply $T \propto N^{5/3}$. The intermediate process of sub-domain folding suggests a more realistic intermediate scaling law of, say,

$$T \propto N^{7/3} \qquad (4)$$

for large proteins. As a matter of fact, a power law dependence of just this form has been proposed[19] on empirical grounds, with the exponent in the range 2.5 to 3. Inserting typical numerical values from Eq. (2), the limiting values of $T$ for a 100 and 1000 amino acid protein are $10^{-3}$ s and 0.3 s respectively. This is comfortably within the maximum time for many protein folds (typically $10^{-6}$ s to $10^{-3}$ s for small proteins in vitro), but near the limit for some, hinting that quantum indeterminism may be a key factor, at least in some cases, of protein folding choreography.

There are many other biological scaling laws, among which the best-known is perhaps the relationship between metabolic rate $P$ and body size $W$:

$$P = aW^\beta \qquad (5)$$

where the scaling exponent $\beta$ is found to be ¾ for mammals and ⅔ for birds[20]. These scaling laws embody the familiar phenomenon that small creatures live fast and die young. Demetrius[21] has attempted to derive the form of Eq. (5), including the exponents, from quantum considerations. He argues that metabolic activity is driven by the production of ATP, which is determined by the transport of both electrons and protons across membranes. He shows that if these processes are constrained by energy quantization,

$$E = n\hbar\omega \qquad (6)$$

then Eq. (5) follows using a plausible model of the bioenergetic cycles involved.

Another general argument concerns the emergence of complexity, of which life is a special case. Seth Lloyd has argued[22] that the inherent randomness of quantum fluctuations constitutes a rich source of algorithmic complexity. He recalls the metaphor of the monkey typing on a typewriter, and observes that there is a much higher probability of the monkey randomly constructing a computer program that will generate complex output states than the probability of those output states being generated *de novo*.

These sorts of argument are at best suggestive. They help delineate the boundaries of the quantum regime and invite further study, but they are no substitute for direct evidence of quantum effects at work in biology. From time to time claims are made that certain specific biological phenomena require non-trivial quantum processes to be explained. While all such claims remain speculative, or peripheral to core biological processes, it is at least useful to list some of them.

## 4. HINTS OF QUANTUM PROCESSES AT WORK IN BIOLOGY

Both experimental and theoretical work offers circumstantial evidence that non-trivial quantum mechanical processes might be at work in biological systems. Some examples are:

1. Mutations. Ever since Crick and Watson elucidated the structure of DNA the possibility has been seriously entertained that mutations might occur as a result of quantum fluctuations, which would serve as a source of random biological information[23]. Proton tunneling can indeed spontaneously alter the structure of nucleotide bases, leading to incorrect pair bonding. McFadden and Al-Khalili have suggested that, in some circumstances, the genetic code should be regarded as a quantum code, so that superpositions of coding states might occur, leading to spontaneous errors in base pairing[24].

2. Enzyme action. Enzymes are proteins that catalyze biochemical reactions but their hugely accelerated reactions rates, with factors as large as $10^{19}$, are difficult to account for by conventional catalytic mechanisms. Evidence that quantum tunneling plays an essential role has been obtained for many enzyme-driven reactions, and it is likely that tunneling is an important factor contributing to the extraordinary efficiency of enzyme catalysis[25].

3. Genetic code. The origin of the genetic code is a major area of study. Patel has argued that the code contains evidence for optimization of a quantum search algorithm[26]. The replication of DNA involves matching up unpaired nucleotide bases on an unzipped DNA strand with complementary free bases from the environment. This is accomplished by a DNA polymerase enzyme that envelops the reaction region, and moves along the strand, sucking in nucleotides at random and forging the links one by one. All else being equal, there is a 0.25 probability that the correct base will be sampled at each stage (there being 4 varieties of bases). Patel points out that a quantum search algorithm will improve the sampling efficiency by a factor of 2. Grover's algorithm in the theory of quantum computation is directed toward searching for a target amid an unsorted database of $N$ objects, and achieves a $\sqrt{N}$ improvement in efficiency over a classical search[27]. If $Q$ is the total number of sampling operations, then Grover's algorithm states

$$(2Q + 1) \arcsin(1/\sqrt{N}) = \pi/2, \tag{7}$$

which has the suggestive solutions

$$Q = 1, N = 4 \tag{8}$$

$$Q = 3, N = 20.2. \tag{9}$$

The universal genetic code is based on triplets of nucleotides of 4 varieties, that code for 20 or 21 amino acids. The fact that these three numbers all come out of Grover's algorithm is either a happy numerical coincidence, or it implies that life uses some form of quantum information processing at the subcellular level. Patel has argued that the nucleotide bases can remain in a quantum superposition for long enough to participate in the replication process[26].

4. Supersymmetry. A different approach to the genetic code is the application of group theory to the coding assignments. This has led to the characterization of the code using supersymmetry[28], a concept borrowed from the unified description of bosons and fermions in particle physics. While this work is mathematically intriguing, there is no obvious physical mechanism for supersymmetry to enter the code via the evolutionary process.

5. Quantum nanostructures. The living cell is a collection of nanomachines that approach the quantum limit. Quantum electrodynamical effects such as the Casimir effect, and quantum vacuum distortion near surfaces such as cell membranes, could become significant. Many structures of interest seem to be located in the quasi-classical realm, and their quantum aspects remain to be established. A good example is the polymerase enzyme that crawls along DNA forging the base pairs. This is a molecular motor, or ratchet, powered by ATP and using nucleotides as the raw material for the base pairing. The physics of this motor is not well understood. For example, what determines the normal operating speed of the motor? Is this limited by quantum processes, or by purely classical considerations such as the density of nucleotides (i.e. the supply of fuel), or chemical kinetics ($kT$)? Let us suppose it is the former. The Wigner inequality (1) may be converted to a velocity bound

$$v > \hbar/mL. \tag{10}$$

The polymerase protein has a mass of about $10^{-19}$ gm, and a length of about $10^{-3}$ cm, yielding a maximum velocity of about $10^{-5}$ cm s$^{-1}$. The experimental data[29] yields a velocity of about 100 base pairs per second, which is indeed about $10^{-5}$ cm s$^{-1}$, suggesting that in normal operation the motor is limited by quantum synchronization uncertainty. Experiments demonstrate that applying tension to DNA using optical tweezers decelerates the motor at a rate of about 3 bases per second per pN of applied tension[29]. At a tension of about 40 pN the motor stops altogether. (With further stretching of the DNA the motor runs backward.) This suggests that the speed of the motor isn't determined by the availability of nucleotides or $kT$ (which don't change as the tension is increased); rather, it depends in some way on the geometrical configuration of the components. The slight deformation caused by the stretching serves to disrupt either the production of ATP or the efficiency of the pair bonding reactions, or both. This suggests that some form of sharply-peaked quantum resonance process, either a tunneling or scattering resonance, might be involved.

How can this hypothesis be tested? Some experiments suggest themselves:

(i) Alter the nucleotide mix and determine the effect on motor speed.
(ii) Deuterate the nucleotides and measure the motor speed.
(iii) Observe the motor in real time using a synchrotron source.

Polymerase is just one of many molecular motors at work in the living cell. Another that has been closely studied for quantum effects is actomyosin. The actin enzyme moves along the myosin filament in muscle cells (it also plays a role in neurons), powered by ATP. This system has been studied by Matsuno[30], who claims that quantum coherence is at work over a long enough range to encompass several adjacent molecules. (He computes a de Broglie wavelength for the motor of 4.5 nm.) Interestingly, the speed of the actin motor ($\sim 10^{-5}$ cm s$^{-1}$) is comparable to that of the polymerase motor.

Another nanostructure of interest is the proton pump, the job of which is to maintain an appropriate voltage across cell membranes[31]. Although these structures are complex enzymes, it may be possible to model their operation as one-dimensional quantum nanotubes. Experiments to measure current flow across the cell membrane may reveal non-trivial quantum effects at work.

Membranes are associated with many complicated surface chemistry and surface physics effects. An early speculation is that the quantization of nonlinear membrane vibrations in cells might exhibit a sort of bose condensate effect[32]. More conventionally, a membrane may be considered as a slab of dielectric (insulator), which polarizes the electromagnetic vacuum in its vicinity. Molecules in this region will suffer changes to their energy levels, and other physical properties, which may be of biological significance[33].

6. Microtubules. Eccles argued that neuron firings are controlled by quantum tunneling processes at the synapses[9]. Hameroff and Penrose have suggested that microtubules inside cells permit long-range quantum coherence, enabling quantum information processing to take place at the sub-cellular level[10]. They use this hypothesis to develop a theory of consciousness.

7. Quantum ratchets, quantum games and quantum cellular automata. There is a burgeoning field of research that seeks to apply insights from engineering to molecular biology and bio-information processing. This includes the theory of ratchets, "order-from-disorder" processes such as Parrondo's games[34], cellular automata and network theory as examples of generating complexity from the iteration of simple rules[35], and game theory[36], in which the emergence of order from disorder in a system (e.g. the evolution of fitness in an organism) may be seen as a "win" in a "game" against nature[37]. All these topics have recently undergone their own "quantum revolution," leading to new classes of phenomena – such as improved payoff possibilities – that have potential biological application[38]. These ideas await decisive experimental confirmation that quantum mechanics is indeed playing a nontrivial role in biological systems.

## 5. THE ORIGIN OF LIFE

Living systems form a very special subset among the set of all complex systems. Biological complexity is distinguished by being *information-based* complexity, and a fundamental challenge to science is to provide an account of how this unique information content and processing machinery of life came into existence[5]. Although discussions of this sort are hampered by the lack of a precise definition of life, most commentators agree that the subset of living systems represents an extremely small fraction of the total space of complex systems[39]. For example, the fraction of peptide chains that have biological efficacy is exponentially small among the set of all possible sequences. Likewise, only a tiny fraction of nucleotide sequences code for biological function; the overwhelming majority of nucleotide sequences would represent biological gobbledygook. Viewed this way, the origin of life is a type of search problem. Given a soup of classical molecular building blocks, how did this mixture "discover" the appropriate extremely improbable combination by chance in a reasonable period of time? Simple calculation shows that it would take much longer than the age of the universe, even if all the matter in the universe consisted of pre-biotic soup, for even a single protein to form by chance[39]. So the classical chance hypothesis seems unsatisfactory.

Quantum mechanics may offer a radical alternative, however. Since quantum systems can exist in superpositions of states, searches of sequence space or configuration space may proceed much faster. In effect, a quantum system may "feel out" a vast array of alternatives simultaneously. In some cases, this speed-up factor is exponential[40]. So the question is: Can quantum mechanics fast-track matter to life by "discovering" biologically potent molecular configurations much faster than one might expect using classical estimates? This is the motivation that underlies the quest for a quantum computer; in effect, quantum computation enables information processing to take place in a large number of states in parallel, thus shortcutting the computational resources necessary to process a given amount of information[6]. Is it conceivable that living systems exploit quantum information processing in some way, either to kick-start life, or to assist in its more efficient running? (Needless to say, if there is quantum information processing taking place within living cells today, there is a huge potential for the quantum computation industry to learn a few new tricks.)

Two fundamental problems stand in the way of this conjecture, however. The first is decoherence, which I shall discuss in the next section. The second concerns the implicit teleological aspect of all searches. A search implies a pre-specified target or destination. Though it is easy to believe that quantum superpositions might accelerate the "discovery" of a specific, special, physical state (e.g. the "living" state), there is an element of teleology creeping into this mode of thought. *We* might be familiar with what it takes for a system to be living, but a molecular mixture isn't. The concept of a "target sequence" or "goal" at the end of a search is meaningless for molecules. Nevertheless, a quantum search could speed up the "discovery" of life if there is some way in which the system "knows when it is getting hot," i.e. if there is some sort of feedback that senses the proximity to life, and focuses or canalizes the search toward it. Sometimes it is claimed[41] that "life is built into the laws of physics," i.e. that there is an inbuilt bias or directionality in physical processes that guide them toward "life." Expressed more neutrally, "life" constitutes an attractor in chemical sequence space. But to suppose that such an attractor is conveniently built into the laws of nature is just too contrived to be believable[5].

However, quantum mechanics may permit a non-teleological theory of "fast-tracking" a chemical system to "life." But first it is necessary to have a definition of the target system. The designation "life" is far too vague. A possible alternative is "replicator," as such an entity has a clear definition in terms of physics, namely, a system that produces a copy of itself. (Simple RNA self-replicators, known as ribozymes, have been discovered, though their prebiotic relevance is questionable.) So the question we are faced with is the following. Suppose in the sequence space of a vast chemical mixture there exists a tiny subset of molecules with the property of being replicators. Imagine now a quantum superposition that encompasses this subset. Will the ability to replicate somehow amplify the wave function (hence probability) of the replicators, at the expense of the other states? Would such a phenomenon be a product of normal linear quantum mechanics, or would it require some sort of nonlinear feedback to "bootstrap" the replicators into existence?

McFadden has argued that such amplification is predicted by ordinary quantum mechanics without even the need for feedback or bootstrapping refinements[23]. His essential conjecture is that a self-replicator would trigger a "collapse of the wave function" (i.e. strong decoherence), thus "locking in" the "discovery" of this crucial molecule. The justification for his conjecture is that replicating molecules impact their environment much more strongly than a random molecule in a quantum soup. If quantum mechanics does indeed encourage a chemical mixture to "zero in" on replicator states as conjectured, then a mixture of nucleotides strands might be encouraged to "discover" the self-replicating configuration with quantum search efficacy. A plausible (though technically highly challenging) experimental scenario would be to construct a chemical soup consisting of a dynamic combinatorial library of RNA molecules linked by quantum tunneling. A self-replicating molecule might be one of an exponentially large number of possible structures that could be synthesized, but the limited resources of the soup would make its appearance highly improbable classically. However, quantum tunneling could allow the library to explore all combinatorial possibilities as a quantum superposition. The presence of the possibility of a replicator within the superposition may be sufficient to trigger environment-induced decoherence.

A different approach has been suggested by Aharonov et. al., based on an appeal to quantum post-selection[42]. In the conventional approach to quantum mechanics, one first defines an initial state, then the system is evolved forward in time unitarily, and a projective measurement is made. But one can also discuss the evolution of a quantum system in which both the initial and final states are specified (pre- and post-selection). In this case, there will generally be *no* pure quantum state in the intervening interval that unitarily evolves from the initial to the final state. For example, the initial state may be a spin-*x* eigenstate of an electron, and the final state a spin-*y* eigenstate. This can lead to remarkable results; for example, the suppression or even elimination of decoherence. Perhaps living systems have the property of being post-selective, and thus greatly enhance the probability of the system "discovering" the living state? Indeed, this might even form the basis of a *definition* of the living state, and would inject an element of "teleology without teleology" into the description of living systems.

## 6. DECOHERENCE EVASION

If quantum mechanics is to play a non-trivial role in bio-systems, then some way to sustain quantum coherence at least for biochemically, if not biologically, significant time scales must be found. Without this crucial step, quantum biology is dead. Simplistic calculations of decoherence rates are very discouraging in this respect – in a warm wet environment like a cell, decoherence times look to be exceedingly short[12,43]. But on a second look, the situation is found to be more subtle.

There are basically two ways in which decoherence might be kept at bay for long enough to enable biologically important processes to occur. The first is screening: if the system of interest can be quasi-isolated from the decohering environment then decoherence rates can be sharply reduced. Very little is known about the screening properties of biological molecules. For example, a reaction region enveloped in an enzyme molecule will be partially screened from van der Waals-mediated thermal interactions from the rest of the cell. Similarly, the histone wrapping of DNA might serve to shield coding protons from decoherence. Matsuno has claimed that organisms may exploit thermodynamic gradients by acting as heat engines to drastically reduce the effective temperature of certain molecular complexes[30]. He cites the example of the slow release of energy from ATP molecules at actomyosin complexes, which he claims implies an effective temperature for the actomyosin of a mere $1.6 \times 10^{-3}$ K. In any event, the lesson of high-temperature superconductivity serves to remind us that in complex states of matter, simple "*kT* reasoning" may be misleading.

The second possibility concerns decoherence-free subspaces. In the effort to build a quantum computer, much attention has been given to identifying subspaces of Hilbert space that are unaffected by the coupling of the system to its environment[6]. Paradoxically, when a system couples very strongly to its environment through certain degrees of freedom, it can effectively "freeze" other degrees of freedom by a sort of quantum Zeno effect, enabling coherent superpositions and even entanglement to persist.

A clear example is provided by a double-well one-dimensional potential[34]. A particle placed in the lowest energy state of one well will tunnel back and forth through the intervening barrier, oscillating with a certain frequency. If the particle is placed instead in an excited state of the well, this flip-flop frequency will be different. Thus an initial state consisting of a superposition of lowest energy and excited states will soon evolve into a complicated muddle as the flip-flops get out of phase. However, if the particle is now allowed to interact strongly with an external heat bath, the environment has the effect of forcing the disparate oscillations into synchrony, thereby maintaining a limited form of quantum coherence, not only in spite of, but because of, environmental interactions. Furthermore, if the system is placed in an entangled state of left and right well-locations, this entanglement is also preserved by environmental interaction. The model was developed in the context of neutrino oscillations, but has general applicability[44].

So far these considerations of decoherence-evasion have an element of wishful thinking to them. The situation would be transformed, however, if unexpectedly long decoherence times could be demonstrated experimentally in a biological setting.

## 6. CONCLUSION

The case for quantum biology remains one of "not proven." There are many suggestive experiments and lines of argument indicating that some key biological functions operate close to, or within, the quantum regime. Quantum effects can be both good and bad news for biology. On the downside, it is clear that quantum mechanics imposes limitations on the operation of biological components, but it remains to be seen whether effects such as quantum uncertainty are a serious impediment to biological function. One area in which further research might prove fruitful is to determine whether the biological clocks that regulate molecular choreography come up against the fundamental quantum limits discovered by Wigner. On the upside, the characterization of bio-systems as information processors opens the way to their exploitation of quantum information processing, perhaps resulting in enhanced – even greatly enhanced – abilities. The deep problem of biogenesis – the discovery of a pathway from non-life to life – may be advanced by viewing the origin of life not so much as a chemical process, but as the emergence of a specific, coded, information-processing system from an incoherent molecular milieu. A full account of biogenesis must address this software aspect. Advances in quantum information theory promise to cast this problem in an entirely new light, and could point the way to the decisive breakthrough in explaining one of sciences deepest mysteries.

**Acknowledgements**

I should like to thank Jim Al-Khalili, John Barrow, Nicole Bell, Lloyd Demetrius, Johnjoe McFadden, Anita Goel, Bernd-Olaf Küppers, Chris McKay, Gerard Milburn, Michael Nielsen and Jeff Tollaksen for helpful discussions.